\documentclass[3p,times]{elsarticle}
\usepackage{newtxtext,newtxmath}
\usepackage[T1]{fontenc}
\usepackage{ae,aecompl}
\usepackage{lineno,hyperref}
\usepackage{graphicx}	
\usepackage{amsmath}	
\usepackage{amssymb}	
\usepackage{subfig}
\usepackage[english]{babel}
\usepackage{fixltx2e}
\modulolinenumbers[5]

\journal{Journal of \LaTeX\ Templates}

\begin{document}

\begin{frontmatter}

\title{Dependence of light scattering properties on porosity, size and composition of dust aggregates}
\author{Prithish Halder\corref{author1}}
\author{Parizath Deb Roy\corref{author2}}
\author{Himadri Sekhar Das\corref{author3}}
\cortext[author1] {First author.\\\textit{E-mail address:} prithishh3@gmail.com}
\cortext[author2] {Second author.\\\textit{E-mail address:} pari.hkd@gmail.com}
\cortext[author3] {Corresponding author.\\\textit{E-mail address:} hsdas13@gmail.com, himadri.sekhar.das@aus.ac.in}

\address{Department of Physics, Assam University, Silchar}




\begin{abstract}
In this work, we study the light scattering properties of dust aggregates ($0.7\mu m \lesssim R_c \lesssim 2.0\mu m$) having a wide range of porosity ($\mathcal{P}$ = 0.59 to 0.98). The simulations are executed using the Superposition T-matrix
code with BCCA, BA, BAM1 and BAM2 clusters of varying porosity. We investigate the nature and dependencies of the different scattering parameters on porosity, size and composition of the aggregated particles for wavelengths 0.45 $\mu$m and 0.65 $\mu$m. We find that the scattering parameters are strongly correlated with the porosity of the aggregated structures. Our results indicate that, when the porosity of the aggregates decreases, keeping characteristic radius of the aggregates ($R_c$) same for all structures, there is an enhancement in the negative polarization branch (NPB) which is accompanied by a substantial increase in the anisotropies present in the material. Also at the exact backscattering region, the anisotropies are found to be linearly correlated with the porosity of the aggregated structure. The computational study reveals that, for low absorbing materials ($k \le 0.1$), the negative polarization minimum ($P_{min}$) is strongly correlated with the associated anisotropies. Finally, we put forward a qualitative comparison between our computationally obtained results and some selected data from the Amsterdam Light Scattering Database for both low and high absorbing materials. The experimental results also suggest that an increase in the NPB is always accompanied by an enhancement in the anisotropy at the backscattering region.

\end{abstract}

\begin{keyword}
Light scattering; Aggregates; Porosity; Modeling; T-matrix; Polarization; Depolarization; Anisotropy
\end{keyword}

\end{frontmatter}


\section{Introduction}
The electromagnetic scattering of light by dust particles found in space has been studied widely using remote sensing and laboratory setups. The physical and chemical properties of cosmic dust particles provide useful evidence about the formation of the planetary system. Cosmic dust is a broader category of dust particles found in space which consists of cometary dust, interplanetary dust, interstellar dust and circumstellar dust particles. The interplanetary dust particles collected from the Earth's stratosphere are found to have an irregular and porous structural arrangement \cite{Brownlee 1985,Warren 1994}. Laboratory simulations, microgravity experiments and computer simulations of IDPs, cometary dust and interstellar dust particles have been carried out to infer a better understanding of the properties of dust particles found in space. These earlier studies of cosmic dust reveal the dependencies of phase function and linear polarization on the scattering angle and wavelength for a variety of cosmic dust samples. Polarimetric study of comets and IDP reveal the existence of negative polarization branch in the backscattering region \cite{Hadamcik 2003, Das 2013, DebRoy 2015}. Laboratory experiments for the measurements of light scattering parameters for aerosols and cosmic dust indicate the presence of negative polarization branch in the backscattering region \cite{Hadamcik 2007, Munoz 2012}. To study the position and maximum of linear polarization of asteroids, the theoretical approach and modified T-matrix method were used by Petrov \& Kiselev \cite{Petrov 2018}. Their study showed that the value of phase angle maximum ($\alpha_{\mathrm{max}}$) strongly depends on the refractive index. They found that the increase of imaginary refractive index leads to increase in $\alpha_{\mathrm{max}}$ and $P_{max}$, whereas the increase of the real part of the refractive index decreases $P_{max}$.

Recently we have studied the light scattering properties of moderately large aggregates with a wide variation in porosity $\mathcal{P}$ from 0.57 to 0.98. This study indicates that if cluster size parameter ($X$) is increased by increasing number of monomers ($N$) then compact clusters show more enhanced NPB compared to that of fluffy clusters whereas, an opposite trend is observed if $X$ is increased by changing the wavelength and keeping $N$ constant. We have also studied the correlation among $P_{max}$, $P_{min}$ and $S_{11}$(180\textdegree) with $\mathcal{P}$. A linear relation is found to exist among them \cite{DebRoy 2017}. We utilize this concept  to explore the NPB and its dependencies on various physical parameters. This negative polarization, on the other hand proves the availability of silicate materials in the dust particles \cite{Kimura 2006, Petrova 2004, Das 2008a, Das 2008b}. It has been suggested that near field effect and the constructive interference of multiply scattered waves give rise to the phenomenon of negative polarization in dust particles though there is no direct evidence or reason behind the actual cause of negative polarization \cite{Tishkovets 2004}. However, some of the recent studies revealed the dependence of anisotropy on the scattering angle for atmospheric dust particles \cite{Nousiainen 2012}. Though anisotropy does not show notable changes for most of the particles in the backscattering region, yet for birefringent or for very low absorbing materials there are some significant variations which are worth studying.

In this paper, we investigate the nature and dependencies of the different scattering parameters on porosity, size and composition of the aggregated particles for wavelengths 0.45 $\mu$m and 0.65 $\mu$m. We will also explore how the negative polarization and anisotropy are correlated.

\section{Aggregate dust model}
Cosmic dust particles are composed of grains which are irregular in shape and whose porosity ranges from very high to very low. The interplanetary dust particles (IDPs) collected from Earth's stratosphere have irregular shapes and fluffy structures, even the interstellar dust grains are also considered to be composed of such aggregate structures \cite{Mathis 1989}. To model dust aggregates, porous and composite aggregates of small spheres (monomers) are generated using various aggregation schemes. Aggregates are created using the Monte-Carlo simulation by random hitting and sticking spheres together. If the process allows a single monomer to join a cluster of monomers, the aggregate is called Ballistic Particle Cluster Agglomeration (BPCA). And if the procedure permits clusters of monomers to stick together then the aggregate is called Ballistic Cluster Cluster Agglomeration (BCCA). BCCA and BPCA clusters are highly fluffy, and a lot of studies has been conducted using such clusters. The porosity of BCCA and BPCA clusters is given by $\approx$ 0.98 and $\approx$ 0.87. To extract the variation or dependence of the light scattering parameters of cosmic dust on the porosity of the dust particles, we need to consider a wide range of porosity from high to low. Shen et. al \cite{Shen 2008} developed three different types of cluster growth which have a wide range of porosity, from high porous (fluffy) to low porous (compact), such as Ballistic Agglomeration (BA), Ballistic Agglomeration with one migration (BAM1) and Ballistic Agglomeration with two migrations (BAM2). BA clusters are similar to the Ballistic Particle Cluster Agglomeration (BPCA) having porosity $\approx$ 0.86 and fractal dimension $\approx$ 3.0. BAM1 clusters are moderately porous with porosity $\approx$ 0.75, and BAM2 clusters are the least porous with porosity $\approx$ (0.59--0.64). In Fig. 1, we show structures of BA, BAM1, BAM2 aggregates.

These clusters are composed of $N$ spherical monomers having monomer radius $a_{m}$. The radius of the equal volume sphere is thus given by \cite{Shen 2009},
\begin{equation}
a_{eff} = N^{1/3} a_{m}
\end{equation}
The porosity of the cluster is defined by the equation:
\begin{equation}
\mathcal{P} \equiv 1 - \left(\frac{a_{eff}}{R_{c}}\right)^{3}\
\end{equation} where R$_{c}$ is the characteristic radius of the cluster and it is related to $a_{eff}$ and the filling factor $f$ by the following relation, \(R_{c} \equiv a_{eff}/f^{1/3}\).\\
\\
The \textit{in situ} measurements of comet Halley showed the presence of magnesium-rich silicates and carbonaceous materials \cite{Jessbeger 1988}, \cite{Jessbeger 1999}. Even interplanetary dust particles are also composed of these materials \cite{Brownlee 1985}. Organic materials, amorphous and crystalline silicate materials are found in comets and IDPs \cite{Rietmeijer 2008}. Amorphous silicates are also found in interstellar and circumstellar medium and constitute 2/3 of the mass of the interstellar dust. A certain fraction of carbonaceous material is found in the interstellar grains\cite{Draine 2003}. The results obtained from the \emph{Rosetta} mission on comet 67P/Churyumov-Gerasimenko indicates the existence of both silicate and organic materials \cite{Schulz 2015, Capaccioni 2015,Goesmann 2015}. Thus silicates and carbonaceous compounds are the main material constituents of comet dust.
In this study, we have considered two compositions, (i) amorphous forsterite (low absorbing) and (ii) amorphous carbon (high absorbing) which are most abundant in cosmic dust. The refractive indices for amorphous silicate are taken from Scott \& Duley (1996) \cite{Scott 1996} having $(n,k)$ = $(1.689, 0.0031)$ and $(1.677, 0.0044)$ at $\lambda$ = 0.45 $\mu$m and 0.65 $\mu$m respectively whereas the refractive indices for amorphous carbon are taken from Jenniskens (1993) \cite{Jenniskens 1993} having $(n,k)$ = $(1.813, 0.479)$ and $(1.93, 0.367)$ at $\lambda$ = 0.45 $\mu$m and 0.65 $\mu$m respectively. These values were already used by Deb Roy et al. \cite{DebRoy 2017} and Kolokolova et al. \cite{Kolokolova 2015} in their dust modelling. The structural parameters used in this study are shown in Table 1.

\begin{figure*}
    \centering
	\includegraphics[width=160mm]{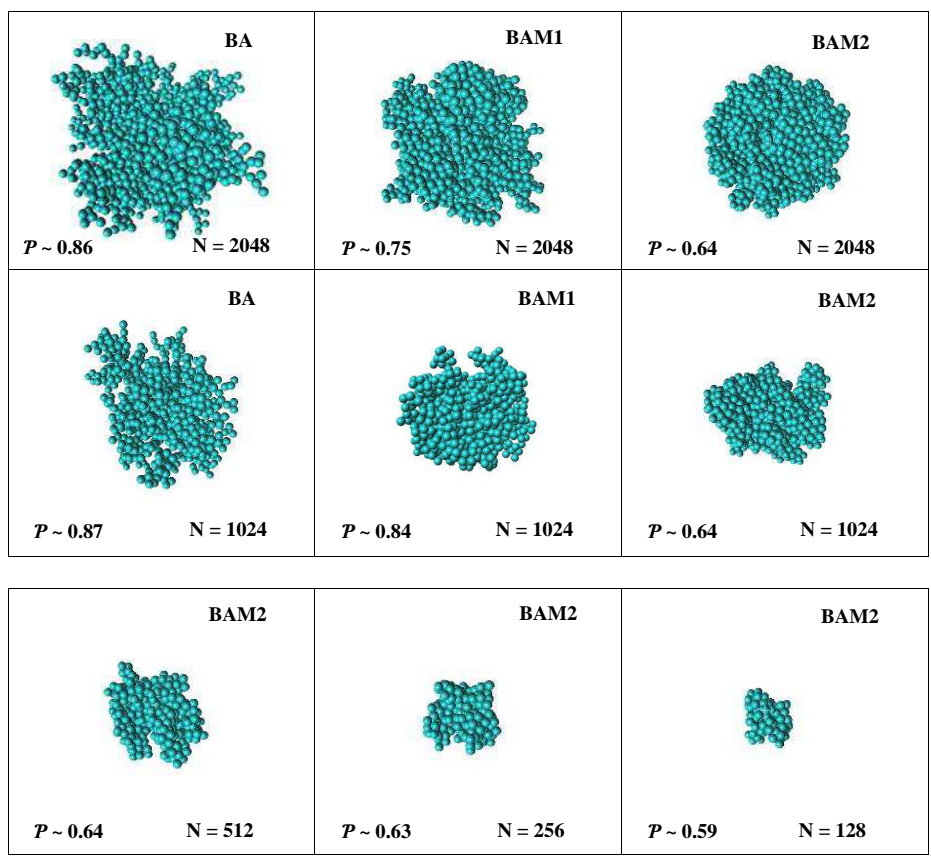}
    \vspace*{-8cm}
    \caption{BA, BAM1 and BAM2 clusters with varying number of monomers ($N$) and porosities ($\mathcal{P}$) are shown (see Table-1 for cluster details). }
    \label{fig:example_figure}
\end{figure*}

\begin{table*}
\caption{Cluster details: Types of aggregate, number of monomers ($N$), structure number as mentioned in Shen et al. \cite{Shen 2008}, porosity of the aggregate ($\mathcal{P}$), ratio of aggregate radius to the effective equal volume radius ($R_{c}/a_{eff}$), monomer radius ($a_m$), effective equal volume radius ($a_{eff}$), characteristic radius of the aggregate ($R_{c}$) and size parameter of the aggregate (\(X_R = 2\pi R_{c}/\lambda\)) at two wavelengths 0.45$\mu$m and 0.65$\mu$m. }
\begin{center}
\begin{tabular}{|c|c|c|c|c|c|c|c|c|c|}
\hline
   Aggregate & $N$ & Structure & $\mathcal{P}$ & $R_{c}/a_{eff}$  & $a_m$ & $a_{eff}$  & $R_{c}$ & $X_R$ & $X_R$ \\
   &       &            \#      &                      &($\mu$m)              &($\mu$m) &($\mu$m) &($\mu$m) & ($\lambda = 0.45\mu m$) & ($\lambda = 0.65\mu m$)\\
 \hline
   BA &  1024 & 1,2,3 & 0.87 & 1.95 & 0.1 & 1.0 & 1.97 & 27.5 & 19.04\\

    &  2048 & 1,2,3 & 0.86 & 1.95 & 0.036 & 0.46 & 0.9 & 12.57 & 8.70 \\
    \hline
     BAM1      &  $1024$ & $1,10,11$ & $0.74$ & $1.58$ & $0.1$ & $1.0$ & $1.58$ & $21.5$ & $15.3$ \\
               &  2048 & 1,8     & 0.75 & 1.61 & 0.044 & 0.56 & 0.9 & 12.57 & 8.70 \\
   \hline
   BAM2 & 128 & 10,12,14 & 0.59 & 1.34 & 0.1 & 0.5 & 0.67 & 9.35 & 6.48 \\
    &  256 & 12,15 & 0.63 & 1.39 & 0.1 & 0.63 & 0.88 & 12.3 & 8.5 \\
    &  512 & 12,15 & 0.64 & 1.41 & 0.1 & 0.8 & 1.13 & 15.78 & 10.9\\
    &  1024 & 1,2,3 & 0.64 & 1.41 & 0.1 & 1.0 & 1.41 & 19.7 & 13.6\\
    &  2048 & 7,8 & 0.64 & 1.41 & 0.050 & 0.64 & 0.9 & 12.57 & 8.70 \\
   \hline
\end{tabular}
\end{center}
\end{table*}


\section{Numerical Simulations}
The numerical calculations for this study are performed using multi-sphere T-matrix code (MSTM) \cite{Mackowski 2011} which gives exact solutions for an aggregate of homogeneous spheres. The MSTM code runs on both serial and distributed memory parallel processing systems. It uses memory and processor resources of the machine and allows for a wide range of calculations and output options without modification and recompilation of the code. MSTM uses message passing interface (MPI) commands to execute on distributed memory multiple processor high-performance computing facility. In these simulations, we have used the T-matrix code for a randomly oriented cluster of spheres.

The connection between the incident and scattered electromagnetic radiation is given by a 4 $\times$ 4 Mueller matrix. This matrix can be defined in a various way depending on the type of normalization. Here we use the scattering matrix S as,

\[
\begin{bmatrix}
    I_{s} \\
    Q_{s} \\
    U_{s} \\
    V_{s}
\end{bmatrix}
=
\frac{1}{k^{2}d^{2}}
\begin{bmatrix}
    S_{11} & S_{12} & S_{13} & S_{14} \\
    S_{21} & S_{22} & S_{23} & S_{24} \\
    S_{31} & S_{32} & S_{33} & S_{34} \\
    S_{41} & S_{42} & S_{43} & S_{44}
\end{bmatrix}
\begin{bmatrix}
    I_{i} \\
    Q_{i} \\
    U_{i} \\
    V_{i}
\end{bmatrix}
\]

where $I$, $Q$, $U$ and $V$ are the Stoke's parameters; $k$ is the wave number and $d$ is the distance from the particle to the observer.
S$_{ij}$ are the orientationally averaged scattering matrix elements. In the present case, the incident light is considered to be unpolarized.

The numerical study presented in this paper revolves around the following five parameters which are defined by the different scattering matrix elements taken from Nousiainen et al. \cite{Nousiainen 2012} and Bohren \& Huffman \cite{Bohren 1983}. \\ \\ \\

\begin{enumerate}
\item[\emph{(i)}] \emph{Degree of linear polarization}: The degree of linear polarization is denoted by $DP$ which is a function of the scattering angle $\theta$ and is represented by, \[DP = -S_{12}/S_{11}\].
\item[\emph{(ii)}] \emph{Phase function}: It is represented by $S_{11}$ and the normalization condition for the phase function is  \[\frac{1}{2} \int_{0}^{\pi}S_{11}(\theta)\sin(\theta)d\theta = 1\]

\item[\emph{(iii)}] \emph{Depolarization Ratio}: This parameter is represented by $D_{1}$ and arises from the anisotropy relation \(S_{11} \neq S_{22}\), where $S_{11}$ and $S_{22}$ are the orientationally averaged scattering matrix elements. Depolarization ratio is calculated as, \[D_{1} = 1 - \frac{S_{22}}{S_{11}}\]
\item[\emph{(iv)}] \emph{Difference parameter}: It is represented by $D_{2}$ and arises from the anisotropy relation \(S_{33} \neq S_{44}\), where $S_{33}$ and $S_{44}$ are the orientationally averaged scattering matrix elements. This parameter is useful to detect birefringent targets. $D_{2}$ is defined by the relation \[D_{2} = \frac{S_{33}}{S_{11}} - \frac{S_{44}}{S_{11}}\].
\item[\emph{(v)}] \emph{Linear Polarization Ratio}: This parameter is represented by $\mu$L and is defined by the equation, \[\mu_{L} = \frac{S_{11} - S_{22}}{S_{11} + 2S_{12} + S_{22}}\]
\end{enumerate}

\section{Results}

In our computations, we have considered two/three cluster realizations of a particular type of aggregate of homogeneous spheres (BA, BAM1 or BAM2) having the same $\mathcal{P}$ and $R_{c}/a_{eff}$ to reduce the variation in the porosity of the aggregates (see Table 1). The porosity along with its error for BA, BAM1 and BAM2 clusters with $N$ = 128, 256, 512, 1024 and 2048 are given in Shen et al. \cite{Shen 2008} (see Table 2 of that paper). It has been reported by many investigators that low absorbing particles show deeper negative branch of polarization (NPB) whereas high absorbing particles show almost negligible NPB \cite{Kimura 2006, Bertini 2007, Lasue 2009, Shen 2009, DebRoy 2017}. Fig. 2(a) shows the variation in the degree of linear polarization, phase function, depolarization ratio and difference parameter for amorphous silicate and amorphous carbon at wavelengths  0.45 $\mu$m and 0.65 $\mu$m respectively for BA cluster having $N$ = 1024 and $a_{m}$ = 0.1 $\mu$m. Fig. 2(b) shows the same variation for BAM2 cluster having $N$ = 1024 and $a_{m}$ = 0.1 $\mu$m. From the figures, it is clear that the silicate particles show both negative and positive branch of polarization with high depolarization ratio and difference parameter which are not observed in the case of carbon particles. Hence in following sections, the study is mainly concentrated on amorphous silicate particles.

\begin{figure*}%
    \centering
    \vspace{-3.0cm}
    \subfloat[]{{\includegraphics[width=70mm]{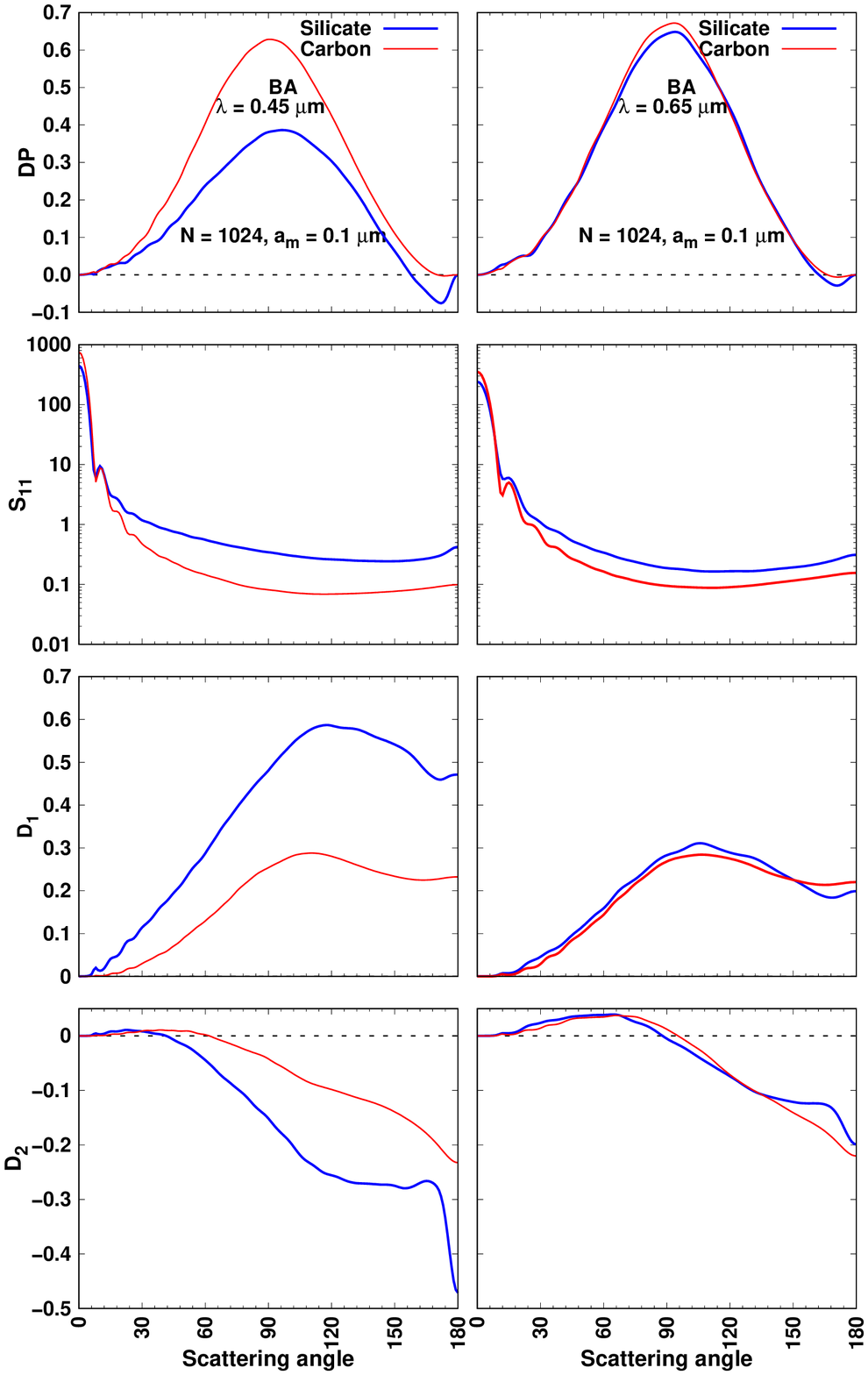} }}%
    \qquad
    \subfloat[]{{\includegraphics[width=70mm]{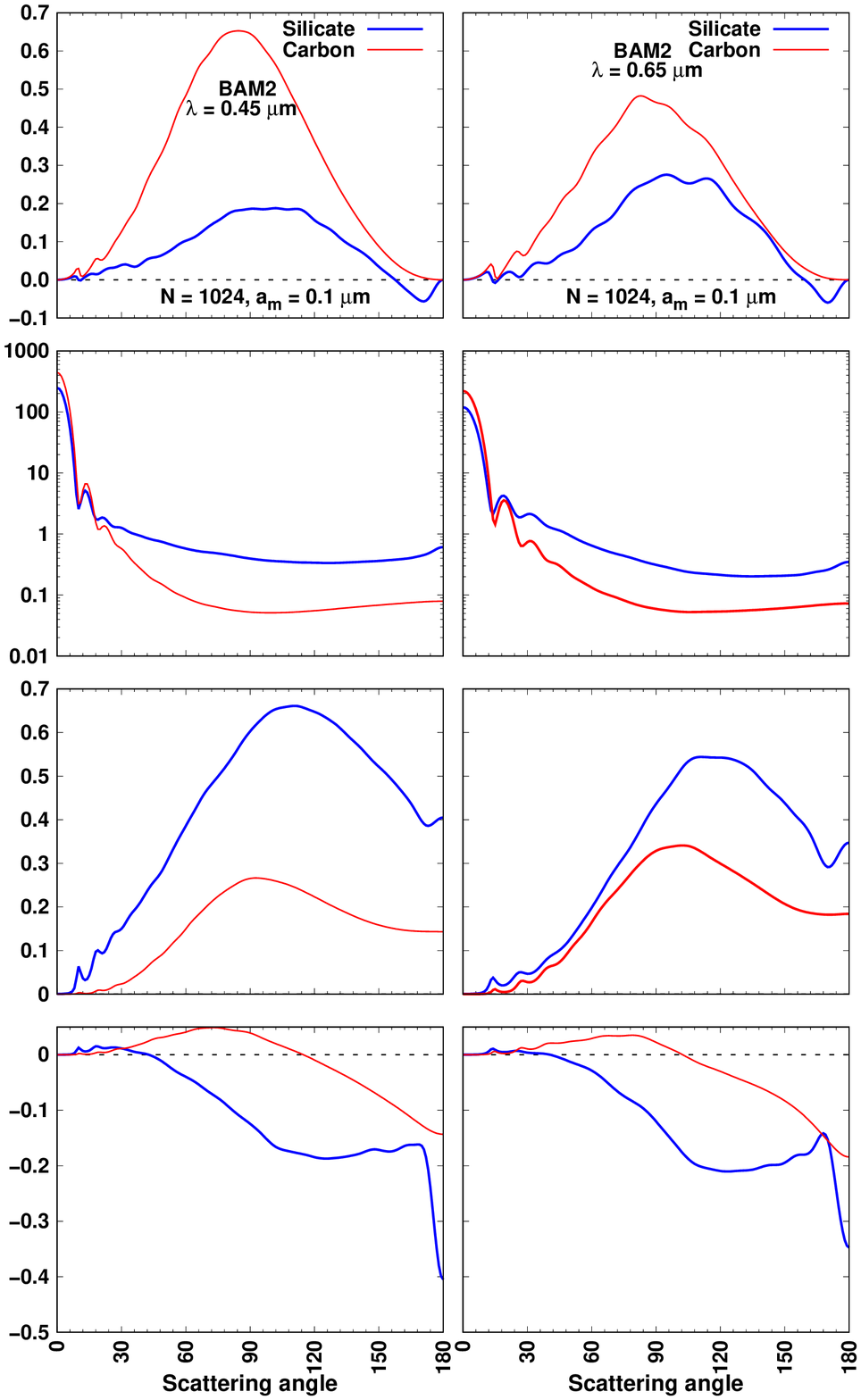} }}%
    \caption{(a) Polarization, phase function, depolarization ratio and difference parameter as functions of scattering angle for amorphous silicate (blue) and amorphous carbon (red) having a$_{m}$ = 0.1 $\mu$m, N = 1024 BA cluster for wavelengths $\lambda$ = 0.45 $\mu$m (left) and 0.65 $\mu$m (right). (b) Same as that of (a), but for BAM2 clusters}.
    \label{fig:example}%
\end{figure*}

\subsection{Effect of porosity}

The porosity of an aggregated structure of dust particles depends on the size, shape, and mixture of grains. With increasing grain size, the porosity of an aggregate shall increase according to equation (2). On the other hand, when $R_{c}$ and $N$ are kept constant, the porosity of the aggregate shall decrease with increasing $a_{m}$. As the scattering cross-section depends on aggregate size, the effect of porosity shall be the dominant feature when $R_{c}$ and $N$ are kept constant. The apparent effect of increasing monomer size is mainly the effect of varying $\mathcal{P}$ or $R_{c}$ as studied by Shen et al. \cite{Shen 2008}. In this section, we study the effect of porosity on different parameters $e.g.$ linear polarization, phase function, depolarization ratio, linear polarization ratio and difference parameter. The light scattering properties of moderately large dust aggregates was studied by our group recently \cite{DebRoy 2017} with a wide range of porosity from 0.57 to 0.98.  We found a strong correlation of $P_{max}$, $P_{min}$ and $S_{11}$(180\textdegree) with porosity ($\mathcal{P}$). It is found that $P_{min}$ and $S_{11}$(180\textdegree) increases linearly with decreasing porosity, whereas $P_{max}$ decreases linearly.

In our calculations we have used BCCA, BA, BAM1 and BAM2 aggregates which have porosities in the decreasing order from high to low. To minimize the effect of aggregates size, we have kept the number of monomers and characteristic size of the aggregates constant. The calculations are carried out for $N$ = 2048 and $R_{c}$ = 0.9 $\mu$m.

As silicate particles show significant degree of negative polarization in the backscattering region, we have studied the effect of porosity for amorphous silicate having refractive index $m = 1.689 \pm i~0.0031$ at $\lambda$ = 0.45 $\mu$m.

In the above calculations, the porosities for BCCA, BA, BAM1 and BAM2 are maintained by keeping $R_{c}$ and $N$ constant and by increasing $a_{m}$. As $R_{c}$ is kept constant the effect of $\mathcal{P}$ becomes dominant over changing monomer size.

\begin{figure*}
    \centering
    \vspace{-8cm}
    \hspace{1.0cm}
	\includegraphics[width=150mm]{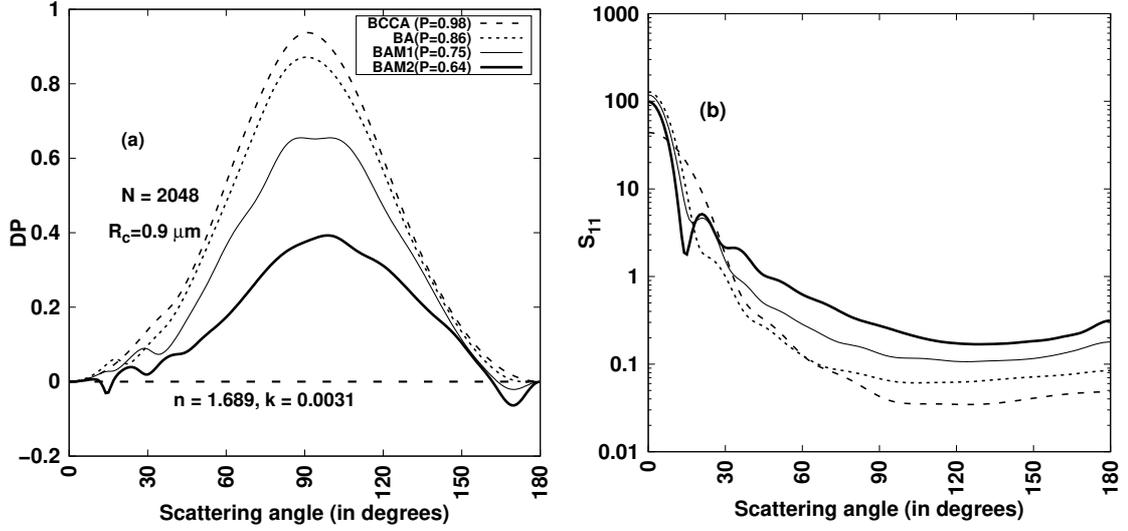}
    \caption{Polarization and phase function as functions of scattering angle for BCCA, BA, BAM1 and BAM2 clusters for amorphous silicate having $N$ = 2048, R$_{c}$ = 0.9 $\mu$m, $\lambda$ = 0.45 $\mu$m  [(a) \& (b)].}
    \label{fig:example_figure}
\end{figure*}

\begin{figure*}
    \centering
    \vspace{-4.0cm}
    \hspace{-3.0cm}
	\includegraphics[width=150mm]{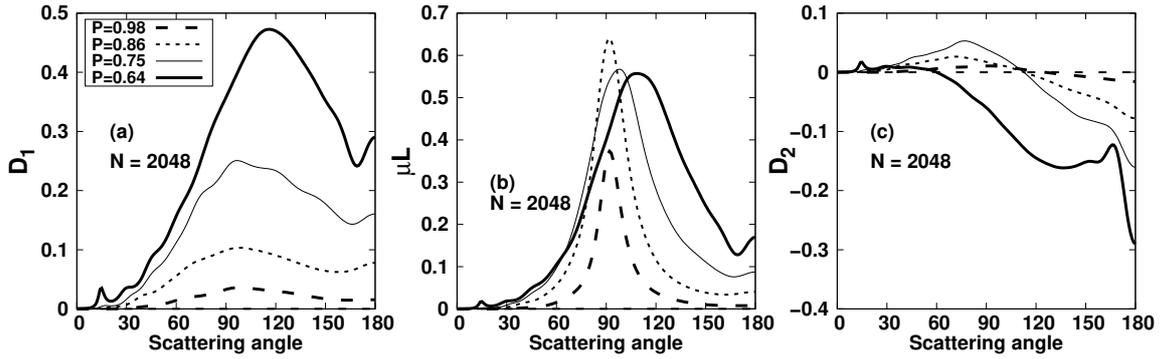}
    \caption{D$_{1}$, $\mu$L and D$_{2}$ as functions of scattering angle for BCCA, BA, BAM1 and BAM2 clusters for amorphous silicate having $N$ = 2048, R$_{c}$ = 0.9 $\mu$m, $\lambda$ = 0.45 $\mu$m [(a), (b) \& (c)]. }
    \label{fig:example_figure}
\end{figure*}

\subsubsection{Effect of $\mathcal{P}$ on $DP$}
Fig. 3(a) depicts the change in linear polarization and phase function due to changing porosity from BCCA $\rightarrow$ BA $\rightarrow$ BAM1 $\rightarrow$ BAM2 for $N$ = 2048. $P_{max}$ shows a decreasing trend with decreasing porosity where particles with high porosity (BCCA) have a high value of polarization,  and particles with low porosity (BAM2) have the lowest value of polarization. Thus we can say, with decreasing porosity the positive branch of polarization decreases significantly.

On the other hand $P_{min}$ intensifies with decreasing porosity, where BA particles show a negligibly small amount of negative polarization (NP), BAM1 particles show a bit intense NP and the highest NP is observed in the case of BAM2 particles. Thus there is a very smooth increase in NP with decreasing porosity. It is to be noted that BCCA particles do not show negative polarization in this case.

The decrease in $P_{max}$ with decreasing porosity can be attributed to the increased electromagnetic interaction among the scatterers as with decreasing porosity the grain grain distance reduces and the coverage of the single wavelength increases which enhances the electromagnetic intensity.

\subsubsection{Effect of $\mathcal{P}$ on S$_{11}$}
Fig. 3(b) shows the change in $S_{11}$ with decreasing porosity for $N$=2048. In this case, a significant increase in S$_{11}$(180\textdegree) is observed with decreasing porosity, where $S_{11}$(180\textdegree) for BCCA is the lowest, and for BAM2 we have the highest value of $S_{11}$. Hence porosity plays a key role in increasing or decreasing the $S_{11}$(180\textdegree). As $S_{11}$(180\textdegree) is directly proportional to the geometric albedo, with decreasing porosity, we must expect a significant increase in \emph{geometric albedo} with decreasing porosity.

\subsubsection{Effect of $\mathcal{P}$ on $D_{1}$}
Fig. 4(a) shows the change in depolarization ratio ($D_{1}$) with decreasing porosity for $N$=2048. In this cases we observe an increase in the depolarization ratio with decreasing porosity where, BCCA particles show minimum $D_{1}$(180\textdegree), and BAM2 particles show maximum $D_{1}$(180\textdegree).

\subsubsection{Effect of $\mathcal{P}$ on $\mu$L}
 Fig. 4(b) shows the change in linear polarization ratio ($\mu$L) with decreasing porosity for $N$=2048. In this case, the linear polarization ratio at 180$^\circ$ ($\mu$L(180\textdegree)) increases with decreasing porosity.   In the backscattering region, the curves develop a depression with decreasing porosity and the depression is strongest for BAM2 particles.

 \subsubsection{Effect of $\mathcal{P}$ on $D_{2}$}
 Fig. 4(c) shows the change in the difference parameter ($D_{2}$) with decreasing porosity for $N$=2048. The change in $D_{2}$ is non-monotonous in the forward scattering region. But in the backscattering region, the magnitude of $D_{2}$ increases with decreasing porosity. As the porosity decreases a peak develops in the backscattering region and the peak is sharpest for the least porosity i.e., BAM2 particles.

\begin{figure*}
    \centering
    \vspace{-4cm}
    \hspace{-3cm}
	\includegraphics[width=150mm]{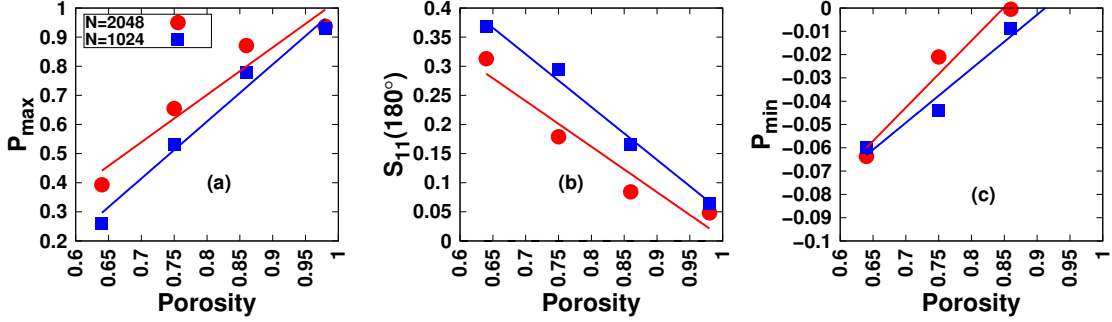}
    \caption{Variation of $(a)$ $P_{max}$ , $(b)$ $S_{11}$ (180\textdegree) and $(c)$ P$_{min}$ with porosity ($\mathcal{P}$) at $\lambda$ = 0.45 $\mu$m. The porosities of BCCA, BA, BAM1 and BAM2 clusters are given by 0.98, 0.86, 0.75 and 0.64, respectively. The \emph{red filled circle} represents the  data for silicate composites for $N$ = 2048, \emph{R$_{c}$} = 0.9 $\mu$m, whereas the \emph{blue filled square} represents the data (taken from Deb Roy et al. (2017) \cite{DebRoy 2017}) for $N = 1024$, R$_c = 1.0$ $\mu$m. The \emph{red} and \emph{blue} straight lines represent the corresponding linear fits.}
    \label{fig:example_figure}
\end{figure*}

\begin{figure*}
    \centering
    \vspace{-4.0cm}
    \hspace{-3.0cm}
	\includegraphics[width=150mm]{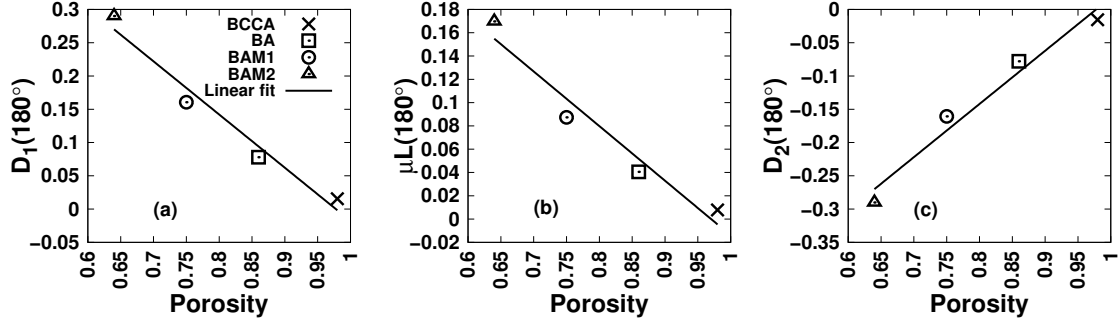}
    \caption{Variation of $(a)$ D$_{1}$(180\textdegree), $(b)$ $\mu$L(180\textdegree) and $(c)$ D$_{2}$(180\textdegree) with porosity ($\mathcal{P}$) of aggregates for $N$ = 2048, \emph{R$_{c}$} = 0.9 $\mu$m at $\lambda$ = 0.45 $\mu$m. The values are taken from Fig.3. The solid lines in $(a)$, $(b)$ and $(c)$ indicate best fit lines.}
    \label{fig:example_figure}
\end{figure*}

\begin{figure*}
    \centering
    \vspace{-4.0cm}
    \hspace{-3cm}
	\includegraphics[width=150mm,scale=1.0]{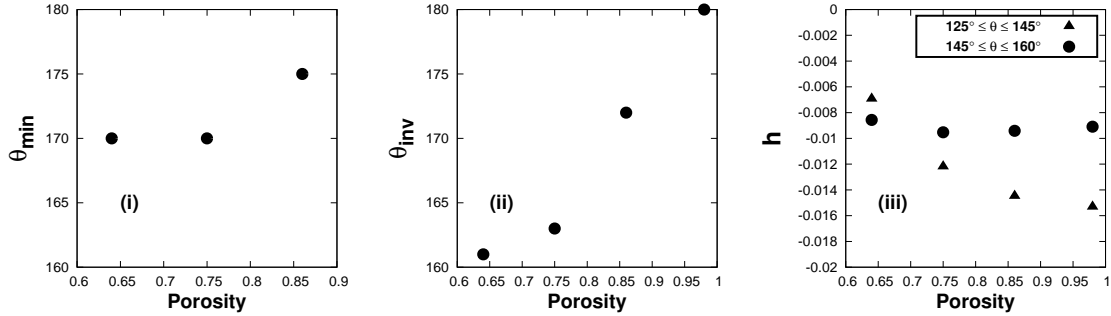}
    \caption{Variation of $(i)$ scattering angle of the negative polarization minimum ($\theta_{min}$), $(ii)$ angle of inversion ($\theta_{inv}$), and $(iii)$ polarimetric slope ($h = \frac{dp}{d\theta}$) for $125^\circ \le \theta \le 145^\circ$ and $145^\circ \le \theta \le 160^\circ$ with porosity $\mathcal{P}$ of aggregates for $N$ = 2048, \emph{R$_{c}$} = 0.9 $\mu$m at $\lambda$ = 0.45 $\mu$m. The values are taken from Fig.3. The porosity of BCCA, BA, BAM1 and BAM2 clusters are given by 0.98, 0.86, 0.75, and 0.64.}
    \label{fig:example_figure}
\end{figure*}

 \subsubsection{Dependence of light scattering parameters on porosity}
In this subsection we study the dependence of different parameters ($P_{max}$, $S_{11}$, $P_{min}$, $D_{1}$, $\mu$L, $D_{2}$, $\theta_{min}, \theta_{inv}$ and polarimetric slope, $h$) on porosity ($\mathcal{P}$) for the aggregates considered above having \emph{N}=2048 and \emph{R$_{c}$}=0.9 $\mu$m for silicate composites at wavelength \emph{$\lambda$}=0.45 $\mu$m. We compare the variation of $P_{max}$(180\textdegree), $S_{11}$(180\textdegree) and $P_{min}$(180\textdegree) with $\mathcal{P}$ from our results, with those obtained by Deb Roy et al. (2017) for $N = 1024$, R$_c = 1.0$ $\mu$m as shown in Fig. 5. The results are almost identical as in both the cases $P_{max}$(180\textdegree), $S_{11}$(180\textdegree) and $P_{min}$(180\textdegree) are linearly correlated with $\mathcal{P}$. Hence we further try to explore the dependence of $D_{1}$(180\textdegree), $\mu$L(180\textdegree) and $D_{2}$(180\textdegree) on $\mathcal{P}$ as shown in Fig. 6. We find that at the exact backscattering region, $D_{1}$(180\textdegree), $\mu$L(180\textdegree) and $D_{2}$(180\textdegree) decreases linearly with increasing $\mathcal{P}$.

We now study the variation of scattering angle of the negative polarization minimum ($\theta_{min}$), angle of inversion ($\theta_{inv}$), and polarimetric slope ($h = \frac{dp}{d\theta}$) for four considered aggregates (shown in Fig. 3(a)) with porosity ($\mathcal{P}$). Instead of considering the usual polarimetric slope, we give two slopes which correspond to the intervals of scattering angles $125^\circ \le \theta \le 145^\circ$ and $145^\circ \le \theta \le 160^\circ$. The results obtained from Fig. 3 are now plotted in Fig. 7. Since negative polarization is not found for BCCA cluster, three data points are plotted in case of Fig. 7(i). It is observed that $\theta_{min}$ is 170$^\circ$ for BAM2 and BAM1 cluster, but 175$^\circ$ in case of BA. Further, $\theta_{inv}$ increases almost linearly when the porosity of aggregates is increased from 0.64 to 0.98 (see Fig. 7(ii)). The polarimetric slope ($h$) is negative and increases in magnitude with increasing porosity for $125^\circ \le \theta \le 145^\circ$. The slope remains almost same for $145^\circ \le \theta \le 160^\circ$.

Our findings suggest that as the aggregate becomes more compact,  $D_{1}(180^\circ)$ and $D_{2}(180^\circ)$ increase along with the substantial increase in the NPB. Since compact cluster BAM2 shows deep or intense NPB, so we will mainly concentrate on BAM2 cluster for the study of NPB in the following sections.

\begin{figure*}
    \centering
    \vspace{-3.0cm}
	\includegraphics[width=100mm,scale=1.0]{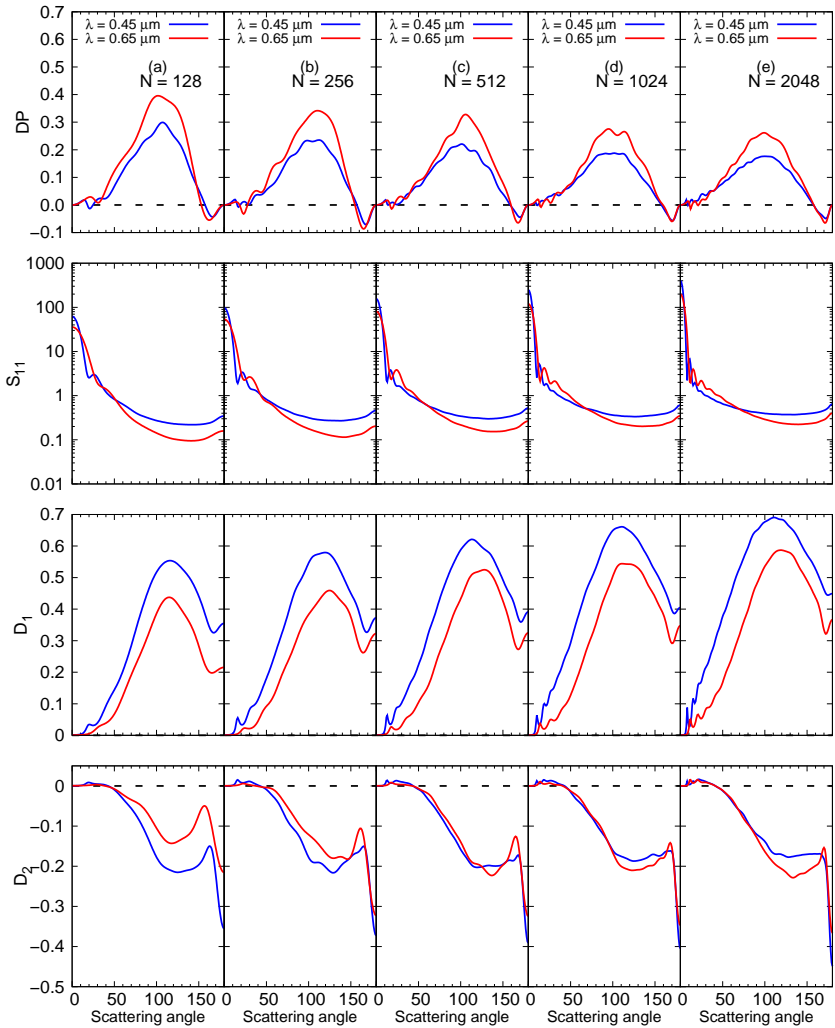}
    \caption{Polarization, phase function, depolarization ratio and difference parameter as functions of scattering angle for amorphous silicate BAM2 clusters of N = 128 (a), 256 (b), 512 (c), 1024 (d) and 2048 (e) respectively having a$_{m}$ = 0.1 $\mu$m for wavelengths $\lambda$ = 0.45 $\mu$m (blue)  and 0.65 $\mu$m (red).}
    \label{fig:example_figure}
\end{figure*}

\begin{figure*}
    \centering
    \vspace{-1.0cm}
    \hspace{-1.5cm}
	\includegraphics[width=100mm,scale=1.0]{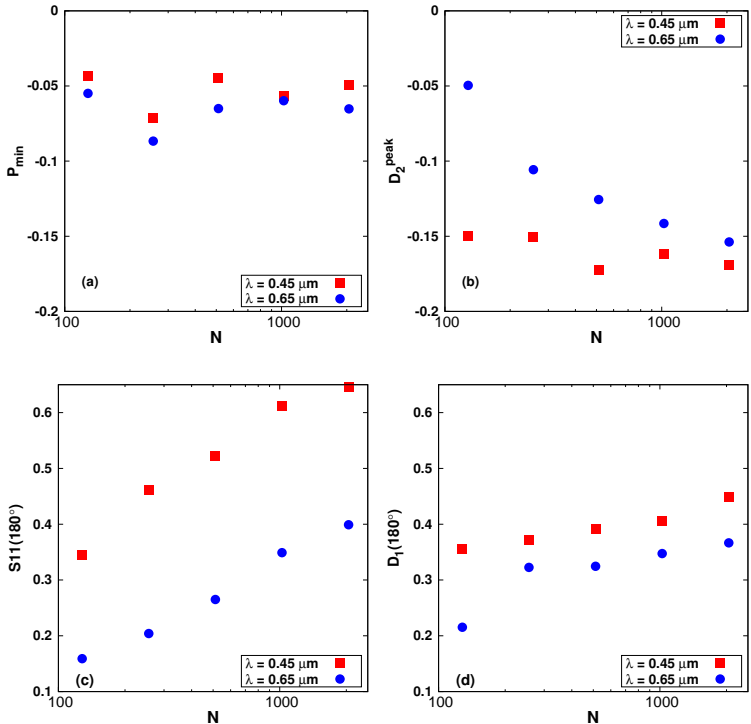}
    \caption{Variation of P$_{min}$, D$_{2}^{peak}$, $S11(180^\circ)$ and $D_1(180^\circ)$ with $N$ for amorphous silicate BAM2 clusters of N = 128, 256, 512, 1024 and 2048 respectively, having a$_{m}$ = 0.1 $\mu$m for wavelengths $\lambda$ = 0.45 $\mu$m  and 0.65 $\mu$m.}
    \label{fig:example_figure}
\end{figure*}

\subsection{Dependence on $N$}
In this section, we study the effect of number of monomers ($N$) on the different light scattering parameters for silicate particles with BAM2 structure at wavelengths $\lambda$ = 0.45 $\mu$m and 0.65 $\mu$m. The monomer size is kept constant at 0.1 $\mu$m and the number of monomers $N$ is varied from 128 $\rightarrow$ 256 $\rightarrow$ 512 $\rightarrow$ 1024 $\rightarrow$ 2048.
Fig. 8 shows the change in linear polarization, phase function, depolarization ratio and difference parameter with increasing number of monomers ($N$) for $\lambda$ = 0.45 $\mu$m and  0.65 $\mu$m respectively due to amorphous silicate particles. The positive branch of polarization decreases with increasing number of monomers ($N$). When $N$ is increased from 128 to 2048, a single wavelength covers more particles which increases the EM interaction among the particles. Hence there is an increase in the scattered intensity which in turn decreases the positive branch of polarization.
This increase in intensity increases the depolarization effect, which shows a significant increasing trend in $D_{2}$.

In Fig. 9, we have shown the dependence of P$_{min}$, D$_{2}^{peak}$, $S11(180^\circ)$ and $D_1(180^\circ)$ on $N$ for amorphous silicate BAM2 clusters of N = 128, 256, 512, 1024 and 2048 respectively, having a$_{m}$ = 0.1 $\mu$m for wavelengths $\lambda$ = 0.45 $\mu$m  and 0.65 $\mu$m. The amplitude of the negative polarization ($P_{min}$) (see Fig. 9(a)) shows a non-monotonous nature with increasing $N$. As the number of monomers are increased, the arrangement of monomers changes, which in turn changes the interference pattern of multiply scattered waves randomly for each set of $N$ and hence there is a non-monotonous nature in NPB \cite{Petrova 2004}. However, the magnitude of $D_2^{peak}$ increases with increase in $N$ (see Fig. 9(b)). It is interesting to observe that $S11(180^\circ)$ increases linearly when $N$ is increased from 128 to 2048 at both the wavelengths (see Fig. 9(c)). Also, $D_1(180^\circ)$ increases almost linearly with increase in $N$ (see Fig. 9(d)).

\subsection{Dependence on $a_m$}
In this case, we have considered a single BAM2 structure having number of monomers $N$ = 2048 for the study of light scattering properties of aggregated dust particles with the variation of monomer radius $a_m$. Fig. 10 shows the change in the degree of linear polarization, phase function, depolarization ratio and difference parameter with increasing monomer radius $a_m$ = 0.03 $\mu$m to 0.11$\mu$m having step size 0.02 $\mu$m for wavelengths $\lambda$ = 0.45 $\mu$m and 0.65 $\mu$m respectively. The positive branch of polarization shows an abrupt decreasing trend with increasing $a_m$. With increasing $a_m$ the scattering cross-section of the aggregate increases due to which the electro-magnetic interaction among the particles in the neighborhood increases. This in turn increases the scattering intensity as shown in Fig. 10 where phase function shows an increasing trend with increasing $a_m$. This increase in phase function increases the depolarization and hence there is a decrease in the positive branch of polarization.

\begin{figure*}
    \centering
    \vspace{-1.0cm}
	\includegraphics[width=100mm,scale=1.0]{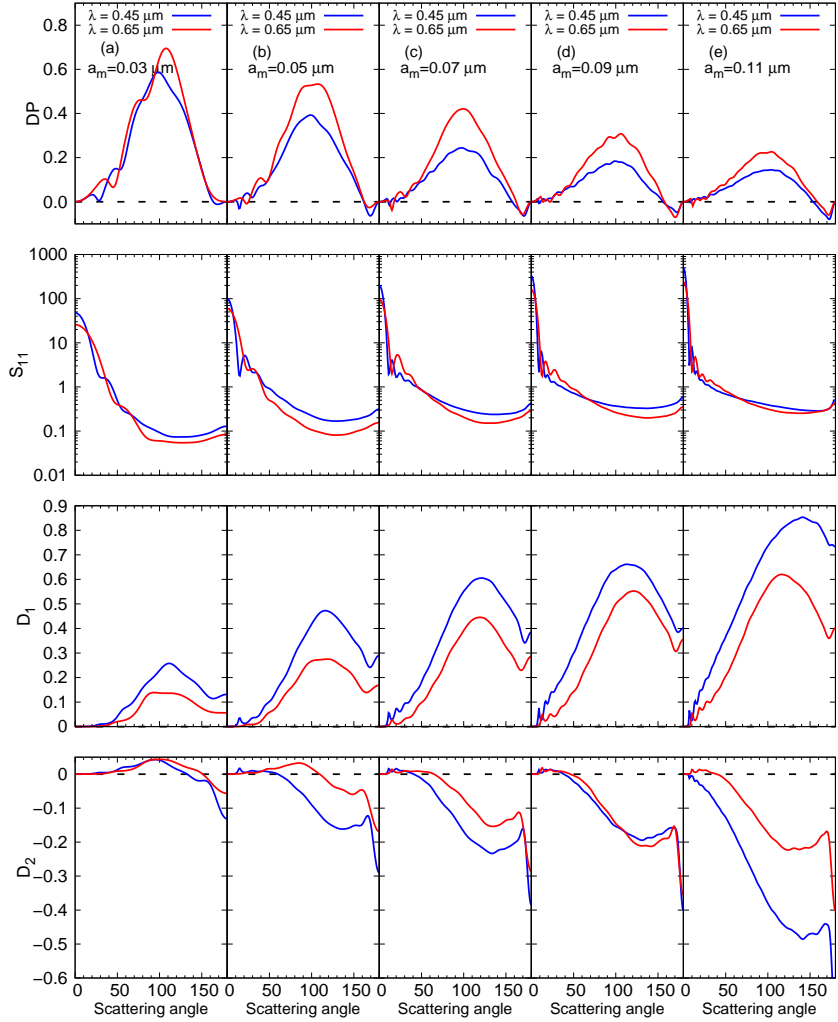}
    \caption{Polarization, phase function, depolarization ratio and difference parameter as functions of scattering angle for amorphous silicate BAM2 cluster of N = 2048 having (a) $a_{m}$ = 0.03 $\mu$m, (b) 0.05 $\mu$m, (c) 0.07 $\mu$m, (d) 0.09 $\mu$m and (e) 0.11 $\mu$m  for $\lambda$ = 0.45 $\mu$m (blue) and 0.65 $\mu$m (red). The refractive indices for amorphous silicate are $(n,k)$ = $(1.689, 0.0031)$ and $(1.677, 0.0044)$ at $\lambda$ = 0.45 $\mu$m and 0.65 $\mu$m respectively.}
    \label{fig:example_figure}
\end{figure*}

We now study the dependence of P$_{min}$, D$_{2}^{peak}$, $S11(180^\circ)$ and $D_1(180^\circ)$ on $a_m$ and is shown in Fig. 11. The magnitude of $P_{min}$ initially increases with increase of $a_m$ and then a slight non-monotonous nature is noticed (see Fig. 11(a)). The magnitude of D$_{2}^{peak}$ gradually decreases with increase in $a_m$ (see Fig. 11(b)). We also find that  $S11(180^\circ)$ increases with increase in $a_m$ and a strong linear correlation between them is observed (see Fig. 11(c)). Further $D_1(180^\circ)$ is also found to increase with increase in $a_m$ (see Fig. 11(d)).

\begin{figure*}
    \centering
   \vspace{-1.0cm}
    \hspace{-1.5cm}
	\includegraphics[width=100mm,scale=1.0]{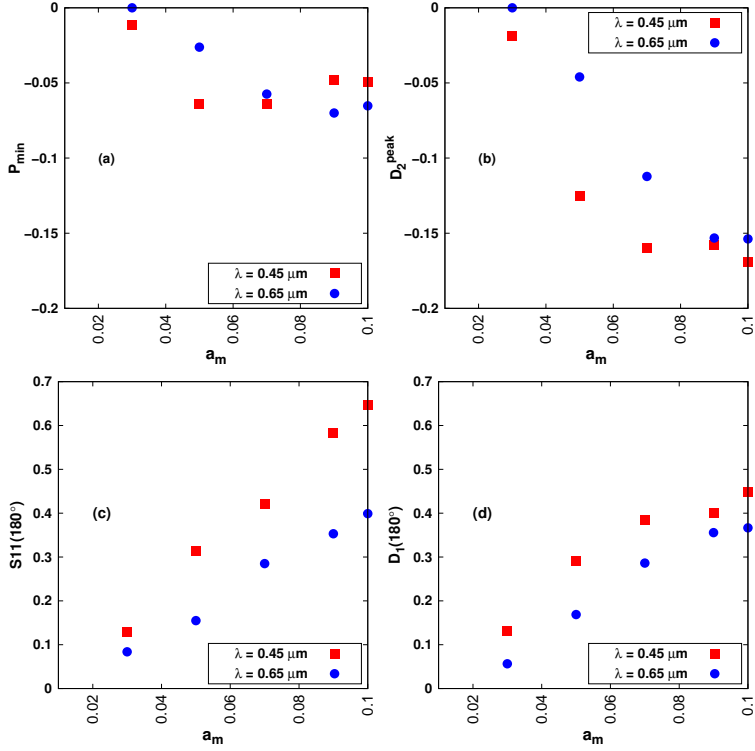}
    \caption{Variation of P$_{min}$, D$_{2}^{peak}$, $S11(180^\circ)$ and $D_1(180^\circ)$ with $a_m$ for amorphous silicate BAM2 clusters of N = 2048 having ce parameter as functions of scattering angle for amorphous silicate BAM2 cluster of N = 2048 having $a_{m}$ = 0.03 $\mu$m, 0.05 $\mu$m, 0.07 $\mu$m, 0.09 $\mu$m and (e) 0.10 $\mu$m  for $\lambda$ = 0.45 $\mu$m (filled circle) and 0.65 $\mu$m (filled square).}
    \label{fig:example_figure}
\end{figure*}

\subsection{Correlation between $P_{min}$ and $D_2^{peak}$}
In this section, we will study how $P_{min}$ and $D_2^{peak}$ are correlated to each other. We will mainly study the effects on negative polarization and anisotropy due to variation in complex refractive indices (`$n$' and `$k$') for $\lambda$ = 0.45 and 0.65 $\mu$m respectively having monomer size \emph{$a_{m}$} = 0.1 $\mu$m and $N$ = 256 BAM2 cluster. Fig. 12 shows the variation of $P_{min}$ and $D_{2}^{peak}$ with increasing values of $n$ for $k$ = 0.001, $N$ = 256 and $a_{m}$ = 0.1 $\mu$m. The variation of $P_{min}$ is very small when $n$ is increased from 1.5 to 1.7. On the other hand the values of $D_{2}^{peak}$ remains close to that of $P_{min}$ for $n$ $\leq$ 1.7 and beyond $n$ = 1.7 the values start to deviate. The real part of the refractive index for amorphous silicate used in the previous sections exist between $n$ = 1.6 to 1.7, and the imaginary part of the refractive index $k$ lies between 0.001 to 0.01. Hence we study the effect of $k$ on the same cluster having same monomer radius for $n$ = 1.6, 1.65 and 1.7. Fig. 13 shows the variation of $DP$, $S_{11}$, $D_{1}$ and $D_{2}$ with increasing $k$ from 0.001 to 0.1 for $n$ = 1.6, 1.65 and 1.7 respectively. When $k$ is increased from 0.001 to 0.1, $P_{max}$ increases, whereas the magnitudes of $S_{11}$(180\textdegree), $D_{1}$(180\textdegree) and $D_{2}$(180\textdegree) decrease. On the other hand, when $n$ increased from 1.6 to 1.7, there is a substantial decrease in $P_{max}$, notable increase of  $D_{1}$(180\textdegree), and $D_{2}$(180\textdegree) and negligible change in $S_{11}$(180\textdegree). Finally, Fig. 14 depicts the effect on
$P_{min}$ and $D_{2}^{peak}$ due to varying $n$ and $k$ for $\lambda$ = 0.45 $\mu$m and 0.65 $\mu$m. We observe that the magnitude of $P_{min}$ and $D_{2}^{peak}$ remains constant at $\lambda$ = 0.45 $\mu$m when $k \le  0.01$ and then starts decreasing. However the magnitude of $D_{2}^{peak}$ remains almost constant for all values of $k$ at $\lambda$ = 0.65 $\mu$m. Thus the real part of the refractive index plays a crucial role in enhancing the anisotropy as well as the negative polarization in the backscattering region.

\begin{figure*}
    \centering
    \vspace{-6.0cm}
	\includegraphics[width=160mm,scale=1.0]{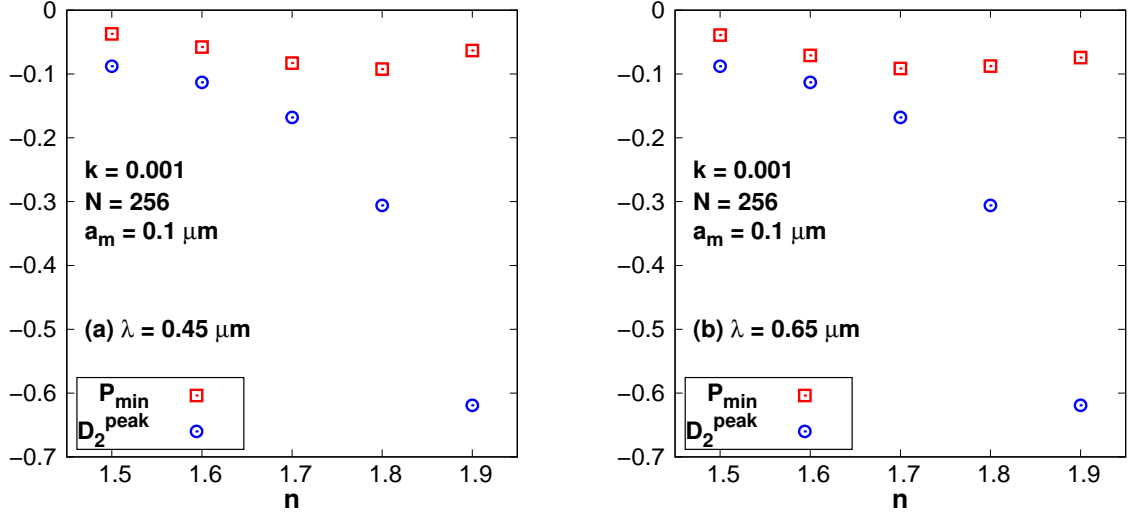}
    \caption{Variation of P$_{min}$ \& D$_{2}^{peak}$ with $n$ for BAM2 clusters having $k$ = 0.001, $N$ = 256, and $a_{m}$ = 0.1 $\mu$m for $\lambda$ = 0.45 $\mu$m (left) \& 0.65 $\mu$m (right). }
    \label{fig:example_figure}
\end{figure*}


\begin{figure*}
    \centering
    \vspace{1.0cm}
	\includegraphics[width=150mm,scale=1.0]{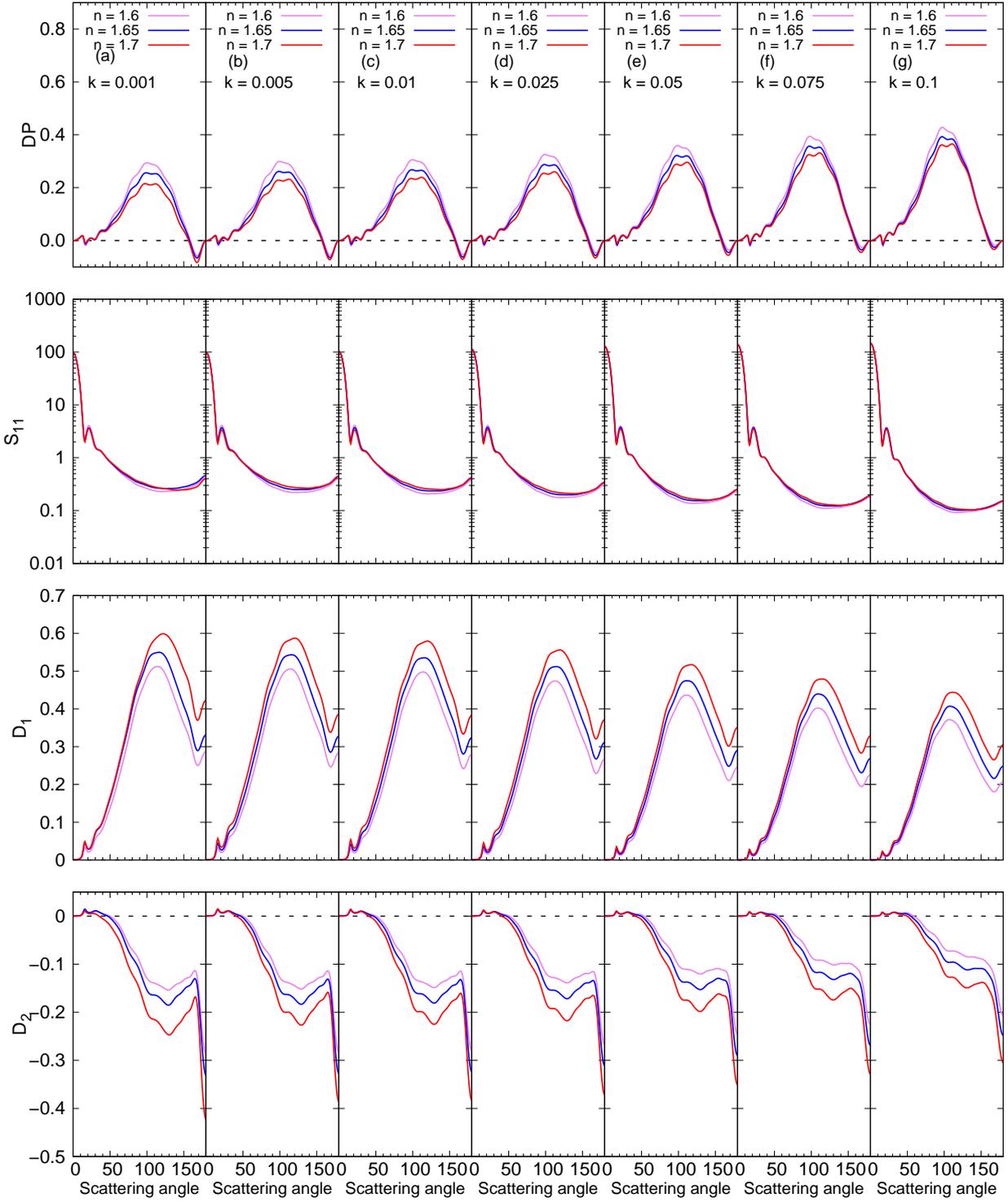}
\vspace{0.5cm}
    \caption{Polarization, phase function, depolarization ratio and difference parameter as functions of scattering angle for amorphous silicate BAM2 cluster of N = 256 and $a_{m}$ = 0.1 $\mu$m for $k$ = 0.001 (a), 0.005 (b), 0.01 (c), 0.025 (d), 0.05 (e), 0.075 (f) and 0.1 (g) having n = 1.6 (pink), 1.65 (blue) and 1.7 (red) for wavelength $\lambda$ = 0.45 $\mu$m.}
    \label{fig:example_figure}
\end{figure*}

\begin{figure*}
    \centering
    \vspace{-0.5cm}
    \hspace{-2.0cm}
	\includegraphics[width=160mm,scale=1.0]{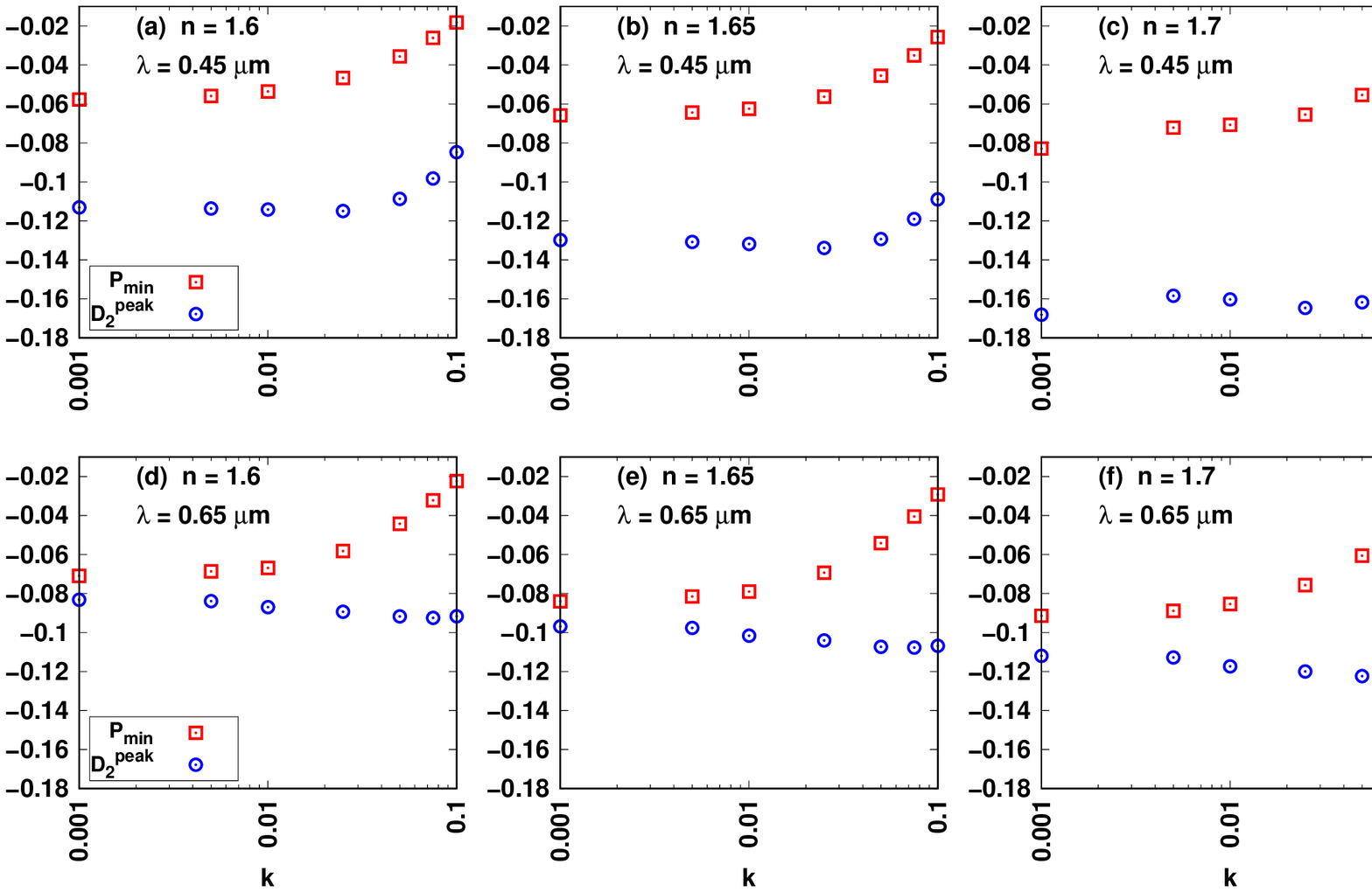}
    \caption{Variation of P$_{min}$ (red square) \& D$_{2}^{peak}$ (blue circle) with $k$ for $n$ = 1.6, 1.65 and 1.7 having N = 256 and $a_{m}$ = 0.1 $\mu$m for $\lambda$ = 0.45 $\mu$m [(a), (b) \& (c)] \& 0.65 $\mu$m [(e), (f) \& (g)] . }
    \label{fig:example_figure}
\end{figure*}

The negative polarization is a unique phenomenon, observed in certain types of materials having imaginary refractive index $k \le 0.1$. Earlier studies claimed that NPB may arise from the internal arrangement of particles where the radiation field is inhomogeneous and the amplitude, phase and the propagation direction changes randomly \cite{Tishkovets 2004}. In amorphous solids or glasses, the phenomenon of birefringence arises from the anisotropy present in the molecular structure and such birefringence is called structural birefringence \cite{Raman 1950}. It is to be noted that $D_{2}$ determines the presence of birefringent material \cite{Nousiainen 2012}. To study how birefringence and negative polarization are correlated to each other, we plot $D_{2}^{peak}$  versus $P_{min}$: $(i)$ keeping $n$ fixed and varying $k$, and $(ii)$ keeping $k$ fixed and varying $n$. The plots are shown in Figs. 15 and 16 for $\lambda$ = 0.45 $\mu$m.
A strong correlation between  $D_{2}^{peak}$ and $P_{min}$ is noticed in both the cases. It can be seen from Fig. 15 that the amplitude of negative polarization ($P_{min}$) decreases with the decrease of $D_2^{peak}$, when $n$ is fixed at some value (1.6, 1.65 or 1.7) and $k$ is increased from 0.001 to 0.1. In Fig. 16, we observe that $D_{2}^{peak}$ and $P_{min}$ can be fitted by a \emph{quadratic regression} equation when $n$ is increased from 1.5 to 1.7 and $k$ is fixed at some value (0.001, 0.0025, 0.005 or 0.0075). It is also important to mention that the change in $D_{2}^{peak}$ is higher in Fig. 16 as compared to Fig. 15. It suggests that the effect of anisotropy is more significant when real part of the refractive index is changing from 1.5 to 1.7 and absorptive index is kept constant at some low values. The similar nature is observed at $\lambda$ = 0.65 $\mu$m (figures are not shown). Thus it is clear from our study that the difference parameter $D_{2}$ (which resembles anisotropy) is affected by the change in real part of the refractive index of the particles ($k \le 0.1$) which in turn affects the negative polarization.

\begin{figure*}
    \centering
    \vspace{-4.5cm}
    \hspace{-3.0cm}
	\includegraphics[width=160mm,scale=1.0]{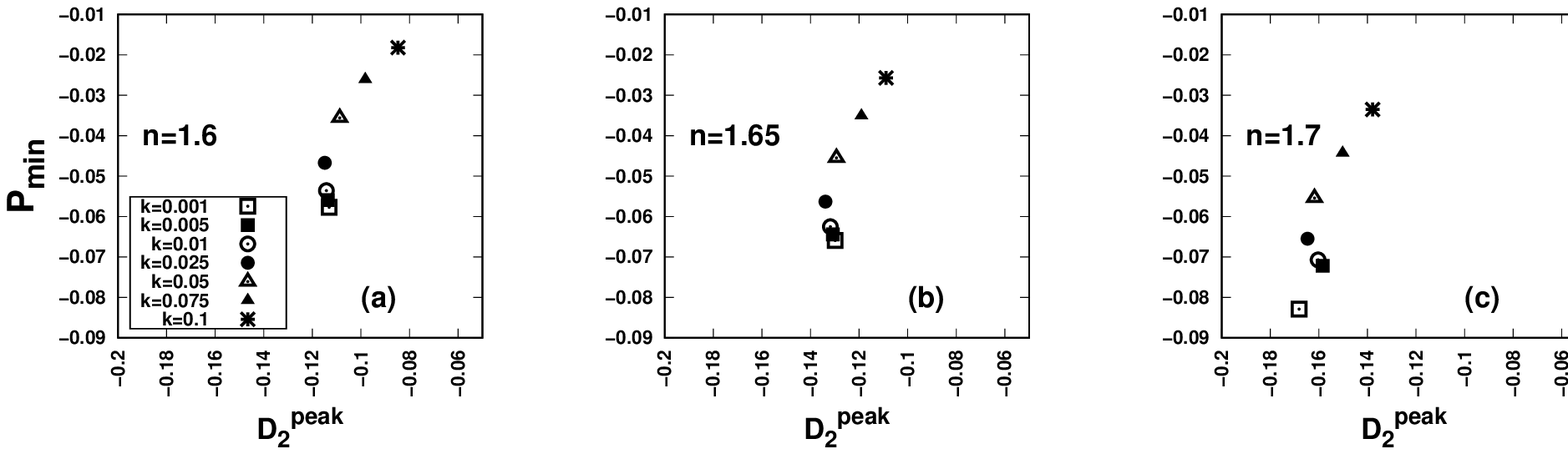}
    \caption{$P_{min}$ is plotted against $D_2^{peak}$ for amorphous silicate BAM2 cluster of N = 256 having  $n = 1.6, 1.65$ and $1.7$ where $0.001 \le k \le 0.1$. The computations have been performed at $\lambda$ = 0.45 $\mu$m.}
    \label{fig:example_figure}
\end{figure*}

\begin{figure*}
    \centering
    \vspace{-4.5cm}
    \hspace{-3.0cm}
	\includegraphics[width=160mm,scale=1.0]{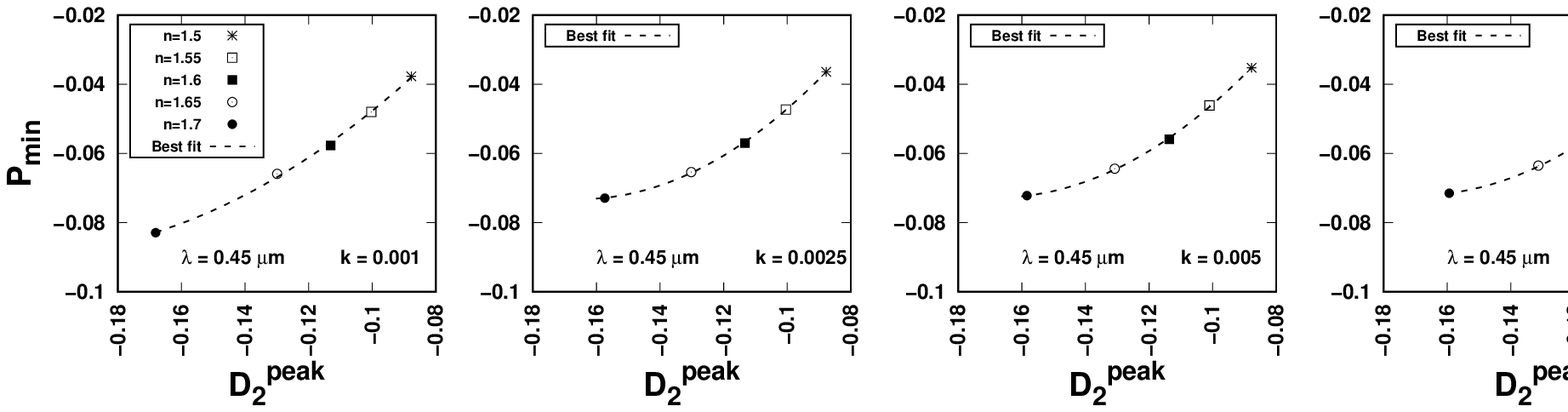}
    \caption{$P_{min}$ is plotted against $D_2^{peak}$ for amorphous silicate BAM2 cluster of N = 256 having  $k = 0.001, 0.0025, 0.005$ and $0.0075$ where $1.5 \le n \le 1.7$. The computations have been performed at $\lambda$ = 0.45 $\mu$m.}
    \label{fig:example_figure}
\end{figure*}

\section{Experimental verification}
In this study, we have mainly used BAM2 clusters to investigate the nature and origin of negative polarization which is observed in cosmic dust particles. To do so, we have considered the difference parameter which is a measure of anisotropy and is used to detect birefringent materials. Our results indicate a dependence of negative polarization on the difference parameter, which is clearly visible in Figs. 15 and 16, where with increasing $D_{2}^{peak}$  there is a significant increase in $P_{min}$ in the backscattering region is observed. Even the sharpness of the peak increases gradually. Again for low absorbing materials both negative polarization, as well as $D_{2}^{peak}$, are very much distinct where as for high absorbing materials both the quantities disappear. In order to verify the above findings, we have considered experimental data of both low and high absorbing materials from the Amsterdam Light Scattering Database \cite{Munoz 2012} to put forward a qualitative comparison between the nature of NPB and $D_{2}^{peak}$ in case of our results and those obtained from the experimental database. Fig. 17(a),(b) represents the variation of the degree of linear polarization and $D_{2}$ for allende \cite{Munoz 2000}, green clay \cite{Munoz 2001}, olivine \cite{Munoz 2000} and forsterite \cite{Volten 2006} from the Amsterdam Light Scattering Database. Table-2 shows the physical and chemical morphologies of the above materials considered in the experiment. Apart from the Amsterdam Light Scattering Database, PROGRA$^{2}$ experiment also provides a large database of linear polarization and phase function for silicates and other materials. As PROGRA$^{2}$ experiment does not provide the scattering matrix elements $S_{22}$, $S_{33}$ and $S_{44}$ which are the essentials elements to calculate the anisotropy, so we have considered only the Amsterdam Light Scattering Database in our study. In the backscattering region, with increasing negative polarization there is an increase in magnitude and sharpness of the $D_{2}^{peak}$, which is in agreement with our findings for low absorbing materials. Fig. 17(c),(d) depict the variation of linear polarization and difference parameter for hematite \cite{Munoz 2006} which is a high absorbing material. In this case we find negligible negative polarization and $D_{2}^{peak}$ which is again in agreement with our findings for high absorbing materials.

\begin{figure*}
    \centering
    \vspace{-1.5cm}
    \hspace{1.0cm}
	\includegraphics[width=150mm]{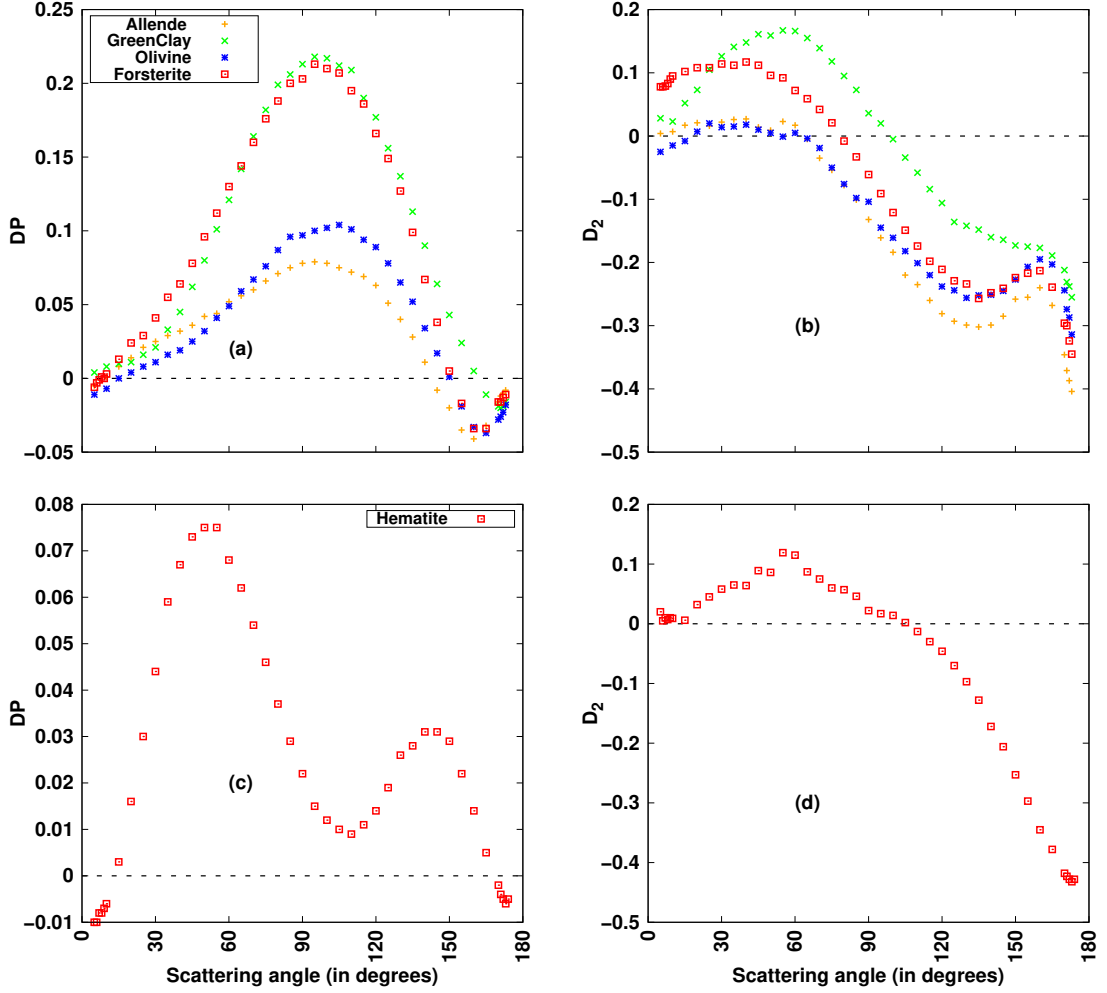}
    \caption{Polarization ($left$) and D$_{2}^{peak}$ ($right$) for low absorbing (allende $[CaAl_{2}Si_{2}O_{8}]$, green clay $[CsAg02]$, olivine $[(Mg,Fe)_{2}SiO_{4}]$ \& forsterite $[Mg_{2}SiO_{4}]$) ($above$) and high absorbing (hematite $[Fe_{2}O_{3}]$) ($below$) materials from the Amesterdam light scattering database}
    \label{fig:example_figure}
\end{figure*}
\begin{table*}[h]
\caption{Material details: Name of the material, chemical formula, refractive index $m$ = n - $i$k, effective radius ($r_{eff}$) and effective variance ($v_{eff}$) at wavelength 0.633$\mu$m taken from Munoz et. al \cite{Munoz 2012}.}
\begin{center}
\begin{tabular}{|c|c|c|c|c|}
\hline
   Material & Chemical formula & $m$ & $r_{eff}$ & $v_{eff}$ \\
   &       &                  &($\mu$m)                   &          \\
 \hline
   Alende &  $[CaAl_{2}Si_{2}O_{8}]$ & 1.65 - $i$0.001 & 0.8 & 3.3\\
    \hline
   Green clay & $[CsAg02]$ & 1.5-1.7 - $i$0.001-0.00001 & 1.55 & 1.8\\
    \hline
   Olivine & $[(Mg,Fe)_{2}SiO_{4}]$ & 1.62 - $i$0.00001 & 1.3 & 1.3\\
    \hline
   Forsterite & $[Mg_{2}SiO_{4}]$ & 1.63 - $i$0.0000 & 1.3 & 3.1\\
    \hline
   Hematite & $[Fe_{2}O_{3}]$ & 3.0 - $i$0.1-0.01 & 0.4 & 0.6\\
   \hline
\end{tabular}
\end{center}
\end{table*}

\section{Summary}

\begin{enumerate}
\item[(i)] The variation in the light scattering parameters $DP$, $S_{11}$, $D_{1}$ and $D_{2}$ is more prominent in case of amorphous silicate as compared to amorphous carbon.

\item[(ii)] The porosity of a material plays a key role in enhancing the negative branch of polarization. For a particular characteristic radius $R_{c}$ (which is $> \lambda$), BCCA cluster shows high $P_{max}$ and low $P_{min}$ whereas BAM2 cluster shows low $P_{max}$ and high $P_{min}$. On the other hand as porosity decreases, the anisotropy inside the material increases which is observed in two parameters $D_{1}$ \& $D_{2}$. The values of $D_{1}$ \& $D_{2}$ are lowest for BCCA cluster and highest for BAM2.

\item[(iii)] At the exact backscattering region, $D_{1}$(180\textdegree), $\mu$L(180\textdegree) and $D_{2}$(180\textdegree) decreases linearly with increasing $\mathcal{P}$, when the characteristic radius of the aggregates ($R_c$) are taken to be same for all structures.

\item[(iv)] $\theta_{min}$, $\theta_{inv}$ and polarimetric slope ($h$) are strongly correlated with the porosity of the aggregates.

\item[(v)] The light scattering parameters show a certain amount of variation with increasing monomer number ($N$). With increasing $N$ from 128 $\rightarrow$ 256 $\rightarrow$ 512 $\rightarrow$ 1024 $\rightarrow$ 2048, the value of $P_{max}$ decreases from 0.30 to 0.18 for $\lambda$ = 0.45 $\mu$m and 0.40 to 0.26 for $\lambda$ = 0.65 $\mu$m. $P_{min}$ shows a non-monotonous nature whereas $S_{11}$ and $D_{1}$ increase with increase in $N$. When $N$ is increased from 128 to 2048, $S_{11}$(180\textdegree) and $D_{2}^{peak}$ increase linearly with $N$. 	

 \item[(vi)] The maximum variation in the light scattering parameters is observed when $a_m$ is increased from 0.03 $\mu$m to 0.11 $\mu$m (keeping $N$ constant at 2048) for BAM2 cluster. With increasing $a_m$, the value of $P_{max}$ decreases from 0.59 to 0.14 for $\lambda$ = 0.45 $\mu$m and 0.69 to 0.22 for $\lambda$ = 0.65 $\mu$m. The phase function at the exact backscattering region $S_{11}$(180\textdegree) increases from 0.13 to 0.51 for $\lambda$ = 0.45 $\mu$m and 0.08 to 0.43 for $\lambda$ = 0.65 $\mu$m. The value of $D_{1}^{max}$ shows an abrupt increase from 0.26 to 0.85 for $\lambda$ = 0.45 $\mu$m and 0.14 to 0.62 for $\lambda$ = 0.65 $\mu$m with increasing $a_{m}$. Also the magnitudes of $D_{2}^{peak}$ or $D_{2}$(180\textdegree) show an increasing trend with increasing $a_m$. Further $D_1(180^\circ)$ is also found to increase with increase in $a_m$.

\item[(vii)] The real part of the refractive index is another parameter which enhances the anisotropy as well as the negative polarization in the backscattering region for low absorbing particles. We have also found that $D_{2}^{peak}$ and $P_{min}$  are strongly correlated to each other when  $n$ is varied from 1.5 to 1.7 and $k$ is fixed at some values (0.001, 0.0025, 0.005 or 0.0075) resembling the low absorbing particles. Our study shows that the difference parameter $D_{2}$ (which resembles anisotropy) is affected by the change in real part of the refractive index of the particles ($k \le 0.1$) which in turn affects the negative polarization.

\item[(viii)] A qualitative comparison between our computationally obtained results and some selected data from the Amsterdam Light Scattering Database for both low and high absorbing materials are made. The experiment results also suggest that an increase in the NPB is always accompanied by an enhancement in the anisotropy at the backscattering region.
\end{enumerate}

\section*{Acknowledgements}
We acknowledge Daniel Mackowski and Michael  Mishchenko, who made their Multi-sphere T-matrix code publicly available. We also acknowledge Bruce T. Draine who made BA, BAM1 and BAM2 clusters publicly available in his website. The anonymous reviewers of this paper are highly acknowledged for their comments and suggestions. This work is supported by the Science and Engineering Research Board (SERB), a statutory body under Department of Science and Technology (DST), Government of India, under Fast Track scheme for Young Scientist (SR/FTP/PS-092/2011). The author P. Deb Roy also wants to acknowledge DST INSPIRE scheme for the fellowship. Part of the computations presented in this publication have been carried out using the High Performance Computing (HPC) facility at IUCAA, Pune. We also acknowledge HPC centre of NIT Silchar in collaboration with C-DAC Pune, where some part of computations were performed.




\begin{thebibliography}{}
\expandafter\ifx\csname url\endcsname\relax
  \def\url#1{\texttt{#1}}\fi
\expandafter\ifx\csname urlprefix\endcsname\relax\def\urlprefix{URL }\fi
\expandafter\ifx\csname href\endcsname\relax
  \def\href#1#2{#2} \def\path#1{#1}\fi

\end{thebibliography}


\begin{thebibliography}{00}


\bibitem {Brownlee 1985}
Brownlee D.E., Cosmic dust: collection and research,  Ann. Rev. Earth Planet. Sci., 13 (1985) 147

\bibitem {Warren 1994}
Warren J.L., Barrett R.A., Dodson A.L., Watts L.A., \& Zolensky M.E., Cosmic Dust Catalog, Nasa Johnson space center, 14 (1994)

\bibitem {Hadamcik 2003}
Hadamcik E., Levasseur-Reourd A.C., Imaging polarimetry of cometary dust: different comets and phase angles, JQSRT, 79-80 (2003) 661-678

\bibitem {Das 2013}
Das H.S., Medhi B.J., Sebastian W., Bertrang G., Deb Roy P. \& Chakraborty A., Polarimetric studies of Comet C/2009 P1 (Garradd), Mon Notice R Astron Soc., 436 (2013) 3500

\bibitem {DebRoy 2015}
Deb Roy P., Das H.S. \& Medhi B.J.,Imaging polarimetry of Comet C/2012 L2 (LINEAR), Icarus, 245 (2015) 241

\bibitem {Hadamcik 2007}
Hadamcik E., Renard J.B., Rietmeijer F.J.M. , Levasseur-Regourd A.C., Hill H.G.M., Karner J.M., Nuth J.A.,
Light scattering by fluffy Mg-Fe-SiO and C mixtures as cometary analogs (PROGRA 2 experiment), Icarus, 190 (2007) 660

\bibitem {Munoz 2012}
Mu\~{n}oz O., Moreno F., Guirado D., Dabrowska D.D., Volten H. \& Hovenier J.W., The Amsterdam-Granada Light Scattering Database, JQSRT, 113 (2012) 565-574

\bibitem {DebRoy 2017}
Deb Roy P., Halder P., Das H.S., Study of light scattering properties of dust aggregates with a wide variation of porosity, Astrophysics Space Sci., 362 (2017) 209

\bibitem {Kimura 2006}
Kimura H., Kolokolova L. \& Mann I., Light scattering by cometary dust numerically simulated with aggregate particles consisting of identical spheres, A \& A., 449 (2006) 1243

\bibitem {Petrova 2004}
Petrova E.V., Tishkovets V.P., Jockers K., Polarization of Light Scattered by Solar System Bodies and the Aggregate Model of Dust Particles, Solar System Res., 38 (2004) 309

\bibitem {Das 2008a}
Das H.S., Das S.R., Paul T., Suklabaidya A. \& Sen A.K., Modelling the polarization properties of Comet 1P/Halley using a mixture of compact and aggregate particles, Mon Notice R Astron Soc., 389 (2008) 787

\bibitem {Das 2008b}
Das H.S., Das S.R. \& Sen A.K., Aggregate dust model to describe polarization properties of Comet Hale-Bopp, Mon Notice R Astron Soc., 390 (2008) 1195

\bibitem {Tishkovets 2004}
Tishkovets V., Litvinov P., Petrova E., Jockers K. \& Mischenko M.: Photopolarimetry in Remote Sensing., Videen G., Yatskiv Y., \& Mishchenko M., Eds. Kluwer Academic Publishers, (2004) 221-242.

\bibitem {Nousiainen 2012}
Nousiainen T., Zubko E., Lindqvist H., Kahnert M., \& Tyynela J., Comparison of scattering by different nonspherical wavelength-scale particles, J. Quant. Spectrosc. Ra., 113 (2012) 2391.

\bibitem {Petrov 2018}
Petrov, D. \& Kiselev, N. Computer simulation of position and maximum of linear polarization of asteroids. J. Quant. Spectrosc. Ra. 2018; 204; 88--93.

\bibitem {Mathis 1989}
Mathis J. S., Rumpl W., \& Nordsieck K. H., Interstellar and cometary dust, Lunar and Planetary Inst., (1989) 48-49

\bibitem {Shen 2008}
Shen Y., Draine B.T. \& Johnson E.T., Modelling porous dust grains with ballistic aggregates. I. Geometry and optical properties, ApJ, 689 (2008) 260

\bibitem {Shen 2009}
Shen Y., Draine B.T. \& Johnson E.T., Modelling porous dust grains with ballistic aggregates. II. Light scattering properties, ApJ, 696 (2009) 2126

\bibitem {Jessbeger 1988}
Jessberger E.K., Christoforidis A. \& Kissel J., Aspects of the major element composition of Halley's dust, Nature., 332 (1988) 691-695

\bibitem {Jessbeger 1999}
Jessberger E.K., Rocky Cometary Particulates: Their Elemental, Isotopic and Mineralogical Ingredients, Space Sci. Rev.,90 (1999) 91-97

\bibitem {Rietmeijer 2008}
Rietmeijer F.J.M, Nakamura T., Tsuchiyama A., Uesugi K., Nakano T. \& Leroux H., Origin and formation of iron silicide phases in the aerogel of the Stardust mission, Meteoritics \& Planetary Science, 43 (2008) 121-134

\bibitem {Draine 2003}
Draine B.T., Scattering by Interstellar Dust Grains. II. X-Rays, ApJ., 598 (2003) 1026

\bibitem {Schulz 2015}
Schulz R., Hilchenbach M., Langevin Y., et al., Comet 67P/Churyumov-Gerasimenko sheds dust coat accumulated over the past four years, Nature, 518 (2015) 216

\bibitem {Capaccioni 2015}
Capaccioni F., Carodini A., Filacchione G., Erard S., Arnold G., et al., Cometary science. The organic-rich surface of comet 67P/Churyumov-Gerasimenko as seen by VIRTIS/Rosetta, Science, 347 (2015) 6220

\bibitem {Goesmann 2015}
Goesmann F., Rosenbauer H., Bredeh$\ddot{o}$ft J.H., Cabane M., Ehrenfreund P., et al., Organic compounds on comet 67P/Churyumov-Gerasimenko revealed by COSAC mass spectrometry, Science, 349 (2015) 6247

\bibitem {Scott 1996}
Scott A., Duley W.W., Ultraviolet and Infrared Refractive Indices of Amorphous Silicates, ApJS, 105 (1996) 401

\bibitem {Jenniskens 1993}
Jenniskens P., Optical constantas of organic refractory residue, A \& A, 274 (1993) 653

\bibitem {Kolokolova 2015}
Kolokolova L., Das H. S., Dubovik O., Lapyonok T., Yang P., Polarization of cosmic dust simulated with the rough spheroid model, PSS, 116 (2015) 30-38

\bibitem {Mackowski 2011}
Mackowski D.W. \& Mischenko M.I., A multiple sphere T-matrix Fortran code for use on parallel computer clusters, J. Quant. Spectrosc. Ra., 112 (2011) 2182

\bibitem {Bohren 1983}
Bohren, C.F. \& Huffman, D.R., Absorption and scattering of light by small particles, John Wiley \& Sons, New York, (1983)

\bibitem {Lasue 2009}
Lasue J., Levasseur-Regourd A.C., Hadamcik E. \& Alcouffe G., Cometary dust properties retrieved from polarization observations: Application to C/1995 O1 Hale Bopp and 1P/Halley, Icarus, 199 (2009) 129-144

\bibitem {Bertini 2007}
Bertini I., Thomas N., Barbieri C., Modeling of the light scattering properties of cometary dust using fractal aggregates, A\&A, 461 (2007) 351

\bibitem {Hanner 1981}
Hanner M.S., Giese R.H., Weiss K., Zerull R., On the definition of albedo and application to irregular particles, A \& A, 104 (1981) 42-46

\bibitem {Raman 1950}
Raman C.V., Structural birefringence in amorphous solids, Proc. Ind. Acad. Sci., 31A (1950) 141

\bibitem {Munoz 2000}
Mu\~{n}oz O., Volten H., de Haan J.F., Vassen W. \& Hovenier J.W., Experimental determination of scattering matrices of olivine and Allende meteorite particles, A \& A, 360 (2000) 777

\bibitem {Munoz 2001}
Mu\~{n}oz O., Volten H., de Haan J.F., Vassen W. \& Hovenier J.W., Experimental determination of scattering matrices of randomly oriented fly ash and clay particles at 442 and 633 nm, Journal of Geophysical Research, 106 (2001) 833

\bibitem {Volten 2006}
Volten H., Mu\~{n}oz O., Brucato J.R., Hovenier J.W., Colangeli L., Waters L.B.F.M. \& van der Zande W.J., Optical and infrared properties of forsterite particles, JQSRT, 100 (2006) 429

\bibitem {Munoz 2006}
Mu\~{n}oz O., Volten H., Hovenier J.W., Min M., Shkuratov Y.G., Jalava J.P., van der Zande W.J. \& Waters L.B.F.M., Experimental and computational study of light scattering by irregular particles with extreme
refractive indices: hematite and rutile, A \& A, 446 (2006) 525



\end{thebibliography}
\end{document}